# Transmission of a Single Plasmon Interacting with Multi-Level Quantum Dots Systems Coupled to Plasmonic Waveguide

Nam-Chol Kim,[†,*] Myong-Chol Ko,[†] Chung-Il Choe[†]

[†]Department of Physics, Kim Il Sung University, Pyongyang, D. P. R. Korea

**Abstract:** We theoretically investigated the transmission properties of a single plasmon interacting with two-level quantum dots (QDs) and a V-type three-level QD, coupled to plasmonic waveguide, respectively. We showed that the transmission and reflection of a single plasmon interacting with multi-level QDs can be switched on or off by controlling the detuning and changing the interparticle distances between the QDs.

**Keywords:** Transmission, Single plasmon, Plasmonic waveguide

[*] Electronic mail: ryongnam19@yahoo.com

Interaction between light and material has always been a central topic in physics, and its most elementary level is just the interaction between a single photon and a single emitter [1]. Manipulating the transmission properties of a single plasmon has attracted particular interests for some fundamental investigations of photon-atom interaction and for its applications in various fields, such as quantum information and quantum devices [2]. The theoretical idea of single-photon transistor has also emerged [3]. Many research groups reported the transmission properties in different quantum systems theoretically and experimentally [4~9]. In the previous studies, they have mainly considered the transmission properties of a single plasmon interacting with one emitter and several emitters, and they mainly focus on the cases where the quantum emitters are all equally spaced each other [9]. In this paper we will investigate the transmission properties of a single plasmon interacting with three non-equally spaced two-level QDs system [Fig. 1(a)] and a two-level QD and a V-type three-level QD system [Fig. 1(b)], coupled to 1D surface plasmonic waveguide, respectively.

We will investigate the transmissioin of a single plasmon interacting with three non-equally spaced two-level QDs (Fig. 1(a)) and a two-level QD and a V-type three-level QD (Fig.1(b)), coupled to 1D surface plasmonic waveguide which is a metal nanowire, respectively and compare the results obtained each other.

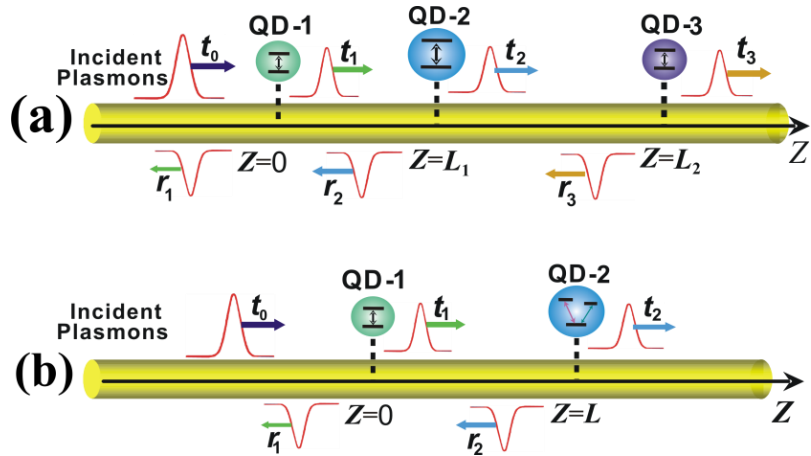

**FIG.1.** (Color online) Schematic diagrams of (a) a system consisting of a single plasmon and three non-equally spaced two-level QDs and (b) a system consisting of a single plasmon and a two-level QD and a V-type three-level QD, coupled to a metal nanowire, respectively. $t_i$ and $r_i$ are the transmission and reflection amplitudes at the place $z_i$, respectively.

By using the real space Hamiltonian and the rotating wave approximation [4], we can solve the equation of the system considered. The Hamiltonian of the coupled system consisting of a single plasmon, a two-level QD and a V-type three-level QD shown in Fig. 1(b) can be written as follows:

$$H = \left(\omega_e^{(2L)} - i\Gamma^{(2L)}/2\right)\sigma_{ee}^{(2L)} + \omega_g^{(2L)}\sigma_{gg}^{(2L)} + \left(\omega_2^{(3L)} - i\Gamma_2^{(3L)}/2\right)\sigma_{22}^{(3L)} + \left(\omega_3^{(3L)} - i\Gamma_3^{(3L)}/2\right)\sigma_{33}^{(3L)} + \omega_1^{(3L)}\sigma_{11}^{(3L)}$$
$$g^{(2L)}\left\{\left[a_r^+(z) + a_l^+(z)\right]\sigma_{ge}^{(2L)} + \left[a_r(z) + a_l(z)\right]\sigma_{ge}^{(2L)}\right\} + g_2^{(3L)}\left\{\left[a_r^+(z) + a_l^+(z)\right]\sigma_{12}^{(3L)} + \left[a_r(z) + a_l(z)\right]\sigma_{21}^{(3L)}\right\}$$
$$+ g_3^{(3L)}\left\{\left[a_r^+(z) + a_l^+(z)\right]\sigma_{13}^{(3L)} + \left[a_r(z) + a_l(z)\right]\sigma_{31}^{(3L)}\right\} + iv_g\int_{-\infty}^{\infty}dz\left[a_l^+(z)\partial_z a_l(z) - a_r^+(z)\partial_z a_r(z)\right]$$

(1)

where $\omega_g^{(2L)}$ and $\omega_e^{(2L)}$ are the eigenfrequencies of the state $|e\rangle$ and $|g\rangle$ of the two-level QD, respectively, and $\omega_j^{(3L)} (j=1,2,3)$ is the eigenfrequency of the state $|j\rangle$ of the three-level QD. The superscripts ($2L$) and ($3L$) correspond to a two-level QD and a V-type three-level QD, respectively. The eigenstate of the system can be constructed as follows,

$$|\psi_k\rangle = \int dz\left[\phi_{k,r}^+(z)a_r^+(z) + \phi_{k,l}^+(z)a_l^+(z)\right]|0,g\rangle_{2L}|0,1\rangle_{3L} + e_k^{(2L)}|0,e\rangle_{2L} + \sum_{j=2}^{3}e_k^{(3L-j)}|0,j\rangle_{3L}, (2)$$

where the kets $|\bullet\rangle_{3L}$ and $|\bullet\rangle_{2L}$ represent the states of a two-level QD and a V-type three-level QD, respectively. $e_k^{(2L)}$ and $e_k^{(3L-j)}$ are the probability amplitudes of the excited state of the two-level QD and the $j$ ($j=2,3$) th excited state of the V-type three-level QD, respectively. The resulting equation of the system can be written as follows ($t_0 = 1$, $r_3 = 0$),

$g^{(2L)}(r_2 e^{ikL} - r_1) - iJ_2^{(2L)}e_k^{(2L)} = 0$, $g^{(2L)}(t_1 e^{-ikL} - t_0) + iJ_2^{(2L)}e_k^{(2L)} = 0$, $g_2^{(3L)}g_3^{(3L)}(r_3 e^{ikL} - r_2) -$

$g_3^{(3L)}J_2^{(3L)}e_k^{(3L-2)} - g_2^{(3L)}J_3^{(3L)}e_k^{(3L-3)} = 0$, $g^{(2L)}(t_0 + r_1) + \Delta_2^{(2L)}e_k^{(2L)} = 0$, $g_2^{(3L)}g_3^{(3L)}(t_2 e^{-ikL} - t_1) +$

$g_3^{(3L)}J_2^{(3L)}e_k^{(3L-2)} + g_2^{(3L)}J_3^{(3L)}e_k^{(3L-3)} = 0$, $g_2^{(3L)}(t_1 + r_2) + \Delta_2^{(3L)}e_k^{(3L-2)} = 0$, $g_3^{(3L)}(t_1 + r_2) +$

$\Delta_3^{(3L)}e_k^{(3L-3)} = 0$, where $\omega_g^{(2L)} = 0$, $\Omega_2^{(2L)} = \omega_e^{(2L)} - \omega_g^{(2L)}$, $\omega_1^{(3L)} = 0$, $\Omega_j^{(3L)} = \omega_j^{(3L)} - \omega_1^{(3L)}$,

$J_2^{(2L)} \equiv g^{(2L)}/v_g$, $J_j^{(3L)} \equiv g_j^{(3L)}/v_g$, $\Delta_2^{(2L)} \equiv i\Gamma_2^{(2L)}/2 - \left(\Omega_2^{(2L)} - \omega_k\right)$,

$\Delta_j^{(3L)} \equiv i\Gamma_j^{(3L)}/2 - \left(\Omega_j^{(3L)} - \omega_k\right)$ ($j=2, 3$). By solving the above set of equations, we can also find the scattering properties of a single plasmon interacting with the system as shown in Fig. 1(b). As for the case shown in Fig. 1(a), one can obtain the solution similarly.

Now, we can compare the scattering properties of a single plasmon by two-level QDs system coupled to 1D surface plasmonic waveguide with the scattering properties of a single plasmon by the V-type three-level QD system coupled to 1D surface plasmonic waveguide. The transmission spectrum of an incident single plasmon by two two-level QDs coupled to 1D surface plasmonic waveguide when the two QDs are placed at the identical position is depicted in Fig. 2 (a), which can also be extracted in Ref [7]. We can also find the transmission spectrum of an incident single plasmon by a V-type three-level QD system coupled to 1D surface plasmonic waveguide [Fig. 2(b)]. What is more interesting is that the transmission spectrum of a single plasmon by a V-type three-level QD is quite the same as that by two two-level QDs placed at the identical position as shown in Figs. 2(a) and 2(b). Finally, we can also compare the scattering properties of a single plasmon by three two-level system with the scattering of a single plasmon by a two-level and a V-type three-level QDs system as shown in Fig. 1(b). We showed the transmission spectra of a single plasmon interacting with three two-level QDs among which the second and third QDs are placed at the identical position as shown in Fig. 2(c). The transmission spectra of a single plasmon interacting with a two-level QD and a V-type three-level QD is quite the same as that with three two-level QDs when the spacing between the two-level QD and the V-type three-level QD, $L$, is equal to the spacing between the first and the second QDs, $L_1$, i. e., $L = L_1$ as shown in Fig. 2(d).

The above results shown in Fig. 2 show that the transmission of a single plasmon in a V-type three-level QD is the same to that in the two two-level QDs with the same transition frequencies, which is because a single incident plasmon can excite only one atomic transition of two two-level QDs or a three-levl QD at a time. Therefore, we can find that one can obtain the same effects of a single plasmon transfer with three-level (or, multi-level) atoms as that with several two-level atoms. As we can see easily from Fig. 2(b), a V-type three-level QD behaves as a perfect mirror at resonance, $\omega_k = \Omega_2^{(3L)}$ or $\omega_k = \Omega_3^{(3L)}$. This occurs because the spontaneous emission which is also part of the guided modes directly results in the reflection. We also find the complete transmission occurs for a V-type three-level QD at $\omega_k = (\Omega_2^{(3L)} + \Omega_3^{(3L)})/2$, at which an incident single plasmon rarely interacts with the V-type three-level QD and thus transmits completely through the QD. The transmission of a single plasmon becomes unity when the incident single plasmons don't interact with QDs.

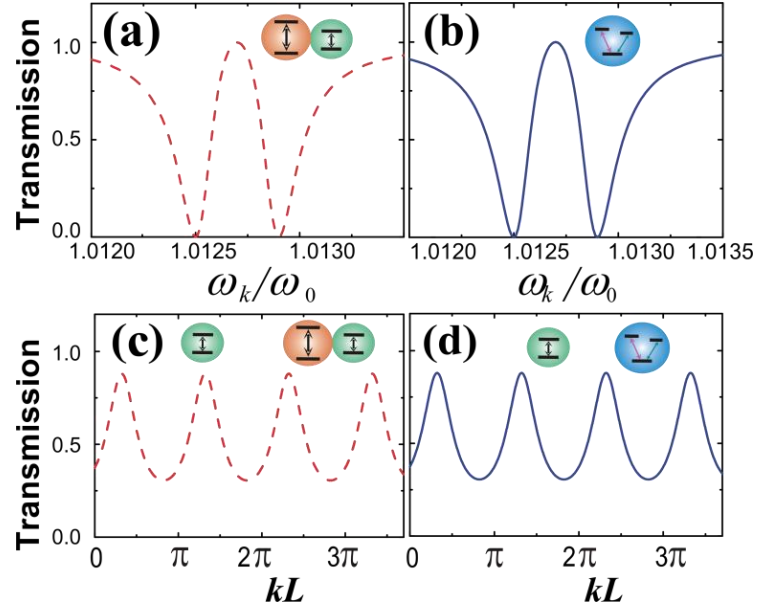

**FIG. 2.** (Color online) Transmission spectra of a single plasmon interacting with two or three-level QDs : (a) two two-level QDs placed at the identical position, (b) a V-type three-level QD, (c) three two-level QDs among which two QDs are placed at the identical position and (d) a two-level QD and a V-type three-level QD. Here we set (a) $\Omega_1=1.0125\omega_0$, $\Omega_2=1.0129\omega_0$, $L=0$, (b) $\Omega_2^{(3L)} = 1.0125\omega_0$, $\Omega_3^{(3L)} = 1.0129\omega_0$, (c) $L_2=0$, $\Delta_k^{(i)}=0.000232\omega_0$, $i = 1, 2, 3$, (d) $\Delta_2^{(2L)}=0.000232\omega_0$, $\Delta_2^{(3L)} = \Delta_3^{(3L)}=0.000232\omega_0$, where the frequency is in units of $\omega_0 \equiv 2\pi v_g / L_1$ and $J_2^{(2L)} = J_2^{(3L)} = J_3^{(3L)} = J = 10^{-4}\omega_0$. Insets show the arrangement of QDs with two-level or three-level energy structures. The stuck QDs represent that they are placed at the identical position.

Several remarks concerned on the experimental realizations should be addressed here. For the realization of the coupling mechanism between a metal nanowire SP and QDs, a silver nanowire and colloidal CdSe/ZnS QDs could be ideal because the exciton energy of CdSe/ZnS QDs is around 2~2.5eV, compatible with the saturation plasma energy (~2.66eV) of the silver nanowire[10]. At the device level, the results studied in this paper can be utilized in such a way that QDs can be attached to a metallic nanowire. In such a scheme, quantum coherence can be generated by an incident laser beam, while the signal is launched through the nanowire as a propagating plasmonon, as shown in Ref. [11]. The propagating plasmons in metal nanowires will finally couple out as photons at the output terminal of the nanowire. In those report, the quantum coherence generated in the QDs by the propagating plasmonon was used to investigate single photon transistor, or control of transport properties of single propagating plsmons. Recently, the first experimental demonstration of plasmon-exciton coupling between silver nanowire (NW) and a pair of quantum dots (QDs) was reported[12], where the separation distances between the two QDs range from microns to 200 nm within the diffraction limit and parameters including the SP propagation length and the wire terminal reflectivity are experimentally determined. Also, single-phton generation, collection, and transport could be successfully accomplished by using one-step scheme which allows for the direct incorporation of single quantum emitters into true 3D photonic structures of nearly arbitrary shape [13].

In summary, we investigated the transmission of a single plasmon by a system including a V-type three-level QD and compared it with that by a system including only two-level QDs. We found that the transmission spectra of a single plasmon interacting with a two-level QD and a V-type three-level QD is quite the same as that with three two-level QDs when the spacing between the two-level QD and the V-type three-level QD is equal to the spacing between the two two-level QDs in three two-level QDs system. Our results may find a variety of applications in the design of the quantum optical devices, such as nanomirrors and quantum switches, and in quantum information processing.

**Acknowledgments.** This work was supported by the National Program on Key Science Research and Key Project for Frontier Research on Quantum Information and Quantum Optics of Ministry of Education of D. P. R of Korea.

**References**

[1] Jiaxiang Zhang, Yongheng Huo, Armando Rastelli, Michael Zopf, Bianca Höfer, Yan Chen, Fei Ding, and Oliver G. Schmidt, Nano Lett. 15 (2015), 422−427.

[2] A. Ridolfo, O. Di Stefano, N. Fina, R. Saija, and S. Savasta, Phys. Rev. Lett.105 (2010) 263601.

[3] D. E. Chang, A. S. Sørensen, E. A. Demler, and M. D. Lukin, Nat. Phys. 3 (2007) 807.

[4] J. -T. Shen, S. Fan, Opt. Lett. 30 (2005) 2001; J. -T. Shen, S. Fan, Phys. Rev. Lett. 95

  (2005) 213001.

[5] M. -T. Cheng, Y. -Q. Luo, P. -Z. Wang, and G. -X. Zhao, Appl. Phys. Lett. 97 (2010) 191903.

[6] M. -T. Cheng, X.-S. Ma, M. -T. Ding, Y. -Q. Luo, and G. -X. Zhao, Phys. Rev. A 85 (2012) 053840.

[7] Nam-Chol Kim, Jian-Bo Li, Zhong-Jian Yang, Zhong-Hua Hao, and Qu-Qian Wang, Appl. Phys. Lett. 97 (2010) 061110; Nam-Chol Kim and Myong-Chol Ko, Plasmonics (Springer) 9 (2014) 9845.

[8] W. Chen, G. -Y. Chen, and Y. -N. Chen, Opt. Express 18 (2010) 10360.

[9] Nam-Chol Kim, Myong-Chol Ko, Qu-Quan Wang, Plasmonics (Springer) 9 (2014) 9846.

[10] A. V. Akimov, A. Mukherjee, C. L. Yu, D. E. Chang, A. S. Zibrov, P. R. Hemmer, H. Park, and M. D. Lukin, Nature (London) 450, 402 (2007).

[11] H. Wei and H. -X. Xu, Nanophotonics, 2 (2013) 155-169.

[12] Qiang Li, Hong Wei and Hongxing Xu, Nano Lett. 14 (2014) 3358-3363.

[13] Andreas W. Schell, Johannes Kaschke, Joachim Fischer, Rico Henze, Janik Wolters, Martin Wegener and Oliver Benson, Scientific Reports. 3 (2013) 1577.